\documentstyle[12pt]{article}
\input epsf 
\begin{document}

\begin{center}
\Large
Axial Magnetostatics of a Ring Current in a Kerr Field \\
A. A. Shatskiy
\end{center}

\begin{center}
Astrospace Center, Lebedev Institute of Physics, \\ 
Russian Academy of Sciences, \\
Profsoyuznaya ul. 84/32, Moscow, 119991 Russia \\
e-mail: shatskiy@lukash.asc.rssi.ru \\
JOURNAL OF EXPERIMENTAL AND THEORETICAL PHYSICS,\\
Vol. 93, No. 5, 2001\\
Received March 28, 2001 \\
\end{center}

Abstract
\begin{center}
The electromagnetic fields generated by a ring current around a Kerr black hole
 have been found.
The acceleration of a charged particle by a force electric field along the rotation
axis is investigated in the constructed
model, as applied to the astrophysics of quasars.
\end{center}

\section{INTRODUCTION}

Studying the interaction of electromagnetic fields
with the gravitational field of a rotating black hole is of
great importance in understanding the astrophysics of
quasars. Quasars manifest themselves as compact,
intense sources of electromagnetic radiation, which
occasionally have huge narrowly directed jets and
which are most likely active galactic nuclei. The Blandford–Znajek 
process [1] is one of the models that
accounts for observational manifestations of quasars. A
magnetohydrodynamic (MHD) model of plasma
accreting onto a rotating black hole underlies this process.
Through the Bardeen–Petterson process [2],
accretion can proceed only from the equatorial plane;
therefore, it makes sense to model accretion as a superposition
of equatorial ring currents. Such modeling is
proper if the pair production by an induced electric field
gives rise to currents that are much weaker than the
source ring current. The Hawking effect of particle production
on the horizon is negligible in this model, because
it gives a negligible correction when the Compton particle
wavelength is much smaller than the radius of space curvature.
For electrons, this corresponds to a black hole with
a mass larger than about $10^{-16}M_{^{_\bigodot}}$.

In all the cited studies, calculations were performed
by using 3 + 1 formalism. In contrast to these studies,
we use the general covariant formalism of general relativity
and do not use the approximation of MHD magnetic
field line freezing-in in plasma, which leads to the
condition for the scalar product of the electric and magnetic
fields being equal to zero (force-free field).
We introduce same notation for electromagnetic and
gravitational quantities as in [3], (\S 90).
\footnote{Below, we use the system of units
$ñ=1$ ---
the speed of light, and
$G=1$ ---
the gravitational constant, for convenience.}

\section{SPECIFYING BOUNDARY CONDITIONS}

Let us specify the boundary conditions that must be
imposed on the electromagnetic-field tensor components
required to determine the latter.
First, the classical boundary conditions must be
satisfied: all field components must become zero at
infinity.
Second, the boundary conditions on the horizon
must also be added to the classical ones. They are
required so that, in the frame of reference associated
with a freely falling observer (FFO), no anomalies arise
in its motion as it flies up to the horizon due to the 
electromagnetic-field acceleration (anomalies result in the
violation of the condition for the FFO electric charge
being a test one.
\footnote{A more detailed discussion of this requirement can be 
found in [4, 5].}
Therefore, to find the conditions
imposed on the field components on the horizon, we
must write out the 4-vector of FFO acceleration and
establish which of its components have the anomalies
on the horizon related to the electromagnetic field.
Below, we use the coordinates that are at rest with
respect to an infinitely remote, static observer.
The FFO 4-acceleration components are
\footnote{Unless stated otherwise, the Latin indices run the 
series 0, 1, 2, 3,
while the Greek indices run the series 1, 2, 3.}
$$
W^i=w^i+\left( {e\over m}\right) F^{ij}u_{_j}
\eqno(1)$$
Here,
$e$
and
$m$
are the FFO test charge and mass, respectively;
$w^i=-\Gamma^i_{km} u^{^k} u^{^m}$,
where $\Gamma^i_{km}$ are the Christoffel symbols;
$F^{ij}$ 
are the contravariant tensor components of
the electromagnetic field; and
$u^{^k}$
are the FFO 4-velocity
components. Let us calculate the 4-acceleration for a
radially falling FFO in the Schwarzschild metric. The
components of the Christoffel symbols and the FFO
4-velocity for this metric are given in [3]. Denoting the
Schwarzschild radius by $r_{_g}=2M$,
where $M$ is the blackhole mass, we have 
$$u^{^0}=-g_{_{rr}}\, ;\quad u^{^r}=-\sqrt{r_{_g}/r}\, 
;\quad
w^0=-{1\over r}(r_{_g}/r)^{3/2}\, 
;\quad w^1={r_{_g}\over 2r^2}(2+g_{_{rr}}).$$
The remaining
$u^{^i}$
and
$w^i$
are zero. When approaching the horizon 
($r\to r_{_g}$), the metric tensor component
$g_{_{rr}}\to -\infty$.
 Therefore, let us write out the asymptotics
of the squares of the observed 4-acceleration components
in the principal (in $g_{_{rr}}$) order expressed in terms
of the contravariant field components:
$$
\begin{array}{lll}
W_0 W^0\to -g_{_{rr}}\left( {e\over m}\right)^2 {r_{_g}\over r}
\left( F^{r0}\right)^2\, ;\quad
W_r W^r\to g_{_{rr}}\left[
{r_{_g}\over 2r^2}\, g_{_{rr}}+\left({e\over m}\right) 
F^{r0}\right]^2 \, ;\\
\makebox{   }\\
W_\theta W^\theta\to g^2_{_{rr}}\left( {e\over m}\right)^2
{r_{_g}\over r}g_{_{\theta\theta}}\left( F^{r\theta}\right)^2\, ;\quad
W_\varphi W^\varphi\to g^2_{_{rr}}\left( {e\over m}\right)^2
{r_{_g}\over r}g_{_{\varphi\varphi}}\left( F^{r\varphi}\right)^2\, .
\end{array}
\eqno(2)$$
We thus see that the singularity in the radial acceleration
component does not result in the violation of the
condition for the charge being a test one for the following
reasons: (1) the gravitation near the horizon in the
radial direction acts a factor of
$g_{_{rr}}$
more strongly than
the electromagnetic field; (2) the zero
$W$
component has
a weak singularity; and (3) the tangential acceleration
components have a strong singularity, which can significantly
change the FFO trajectory near the horizon and
can violate the condition for the FFO charge being a test
one: $e\ll m$.
The contravariant components of the tangential
magnetic field must be set equal to zero on the
horizon, lest this happen. Similarly, the asymptotics of
the squares of the 4-acceleration components dependent
on the covariant field components is
$$
\begin{array}{lll}
W_0 W^0\to -g_{_{rr}}\left( {e\over m}\right)^2 {r_{_g}\over r}
"\left( F_{r0}\right)^2\, ;\quad"
W_r W^r\to g_{_{rr}}\left[
{r_{_g}\over 2r^2}\, g_{_{rr}}+\left({e\over m}\right) 
F_{r0}\right]^2\, ;\\
\makebox{  }\\
W_\theta W^\theta\to g^2_{_{rr}}\left( {e\over m}\right)^2 
g^{^{\theta\theta}}\left( F_{\theta 0}\right)^2\, ;\quad
W_\varphi W^\varphi\to g^2_{_{rr}}\left( {e\over m}\right)^2
g^{^{\varphi\varphi}}\left( F_{\varphi 0}\right)^2\, .
\end{array}
\eqno(3)$$
We thus see that the covariant components of the tangential
electric field with the strongest singularity must
become equal to zero on the horizon. However, the Kerr
field rather than the Schwarzschild field is of physical
interest. Let us write out the Kerr metric and its determinant
in a nonrotating (relative to remote stars) frame:
$$
\left\{
\begin{array}{lll}
ds^2=(1-{r_g r\over\rho^2})dt^2-{\rho^2\over{\scriptstyle\Delta}}\, 
dr^2-
\rho^2\, d\theta^2-\\
-(r^2+a^2+{r_g ra^2\over\rho^2}\sin^2\theta)\sin^2\theta\, 
d\varphi^2
+{2r_g ra\over\rho^2}\sin^2\theta\, d\varphi\, dt\, , \\
-g=\rho^4\sin^2\theta 
\end{array}
\right.
\eqno(4)$$
(here, $\rho^2 =r^2+a^2\cos^2\theta$, 
${\scriptstyle\Delta}=r^2+a^2-r_g r$,
and $a$ is the Kerr parameter). 
In this case, in the Kerr field, both the
space itself and the FFO are drawn into rotation when
approaching a black hole. The conditions for the tangential
electromagnetic field components in the
Schwarzschild field then change to the conditions for
the same components in the frame of reference observers
in the Kerr field.
\footnote{In what follows, by the reference frame we mean a frame in
which $g_{_{0\varphi}}^{''}=0$
(see [4, 5]). This is the frame in which the reference
observers rotate with angular velocity 
$\Omega =-g_{_{0\varphi}}/g_{_{\varphi\varphi}}$
relative
to remote stars and have a zero angular momentum.}
Expressions (2) and (3) give a
summary of the boundary conditions for the electromagnetic
field on the horizon in the Kerr field:
$$
{F''}_{\theta 0}\to 0\, ;\quad {F''}_{\varphi 0}\to 0\, ;\quad
{F''}^{r\theta}\to 0\, ;\quad {F''}^{r\varphi}\to 0\, .
\eqno(5)$$
Let us write out formulas that can be of use in the subsequent
analysis. 
Designating 
$\kappa =-g_{_{00}}/g_{_{\varphi\varphi}}$,
$\kappa +\Omega^2={\scriptstyle\Delta}\sin^2\theta 
/g_{_{\varphi\varphi}}^2$,
we then have
$$
\begin{array}{lll}
g^{^{\theta\theta}}=1/g_{_{\theta\theta}}\, ,\,\,
g^{^{rr}}=1/g_{_{rr}}\, ,\,\,
g^{^{00}}=-{g_{_{\varphi\varphi}}\over{\scriptstyle\Delta}
\sin^2\theta }\, , \\
g^{^{\varphi\varphi}}=-{g_{_{00}}\over{\scriptstyle\Delta}
\sin^2\theta } \, ,\,\,
g^{^{0\varphi}}={g_{_{0\varphi}}\over{\scriptstyle\Delta}
\sin^2\theta } \, .
\end{array}
\eqno(6)$$

\section{PASSAGE TO THE REFERENCE FRAME}

Since the current is axisymmetric and stationary, the
system has two Killing vectors: along the translations
in time and along the translations in angle $\varphi$. Therefore,
all fields in frame (4) are stationary. In general, this 
stationarity
is not conserved when passing to a frame of
reference rotating with an arbitrary angular velocity. To
show this, let us write out the coordinate transformations
to a frame of reference rotating with a spatially
nonuniform angular velocity 
$\omega (r,\theta )$
required in the
subsequent analysis:
$$
dx^i=dx'^k[\delta^i_k+\omega\delta^i_\varphi\delta^0_k +t\delta^i_\varphi
\delta^\beta_k\partial_\beta\omega ]
\eqno(7)$$
The corresponding transformations of the contravariant
field components (see [3]) are
$$
F'^{\alpha 0}=F^{\alpha 0} ;\,\, "
F'^{0\varphi}=F^{0\varphi}+tF^{\alpha 0}\partial_\alpha\omega ;\,\,
F'^{\alpha\varphi}=F^{\alpha\varphi}-\omega F^{\alpha 0} ;\,\,
F'^{r\theta}=F^{r\theta} .
\eqno(8)$$
For the covariant components, we derive 
\footnote{
$e^{\alpha\beta\gamma}=e_{\alpha\beta\gamma}$ --- 
is the Levi–Civita symbol.}
$$
F'_{\alpha\varphi}=F_{\alpha\varphi} ;\,\,
F'_{\alpha 0}=F_{\alpha 0} +\omega F_{\alpha\varphi} ;\,\,
F'_{0\varphi}=F_{0\varphi} ;\,\,
F'_{r\theta}=F_{r\theta}+t\, e^{\alpha\beta\varphi}
F_{\alpha\varphi}\partial_\beta\omega .
\eqno(9)$$
In expressions (8) and (9), $\alpha$ runs the values of $r$ 
and $\theta$.
We see from these expressions that only the frames of
reference rigidly rotating relative to another stationary
frame (relative to remote stars) are stationary. The stationary
frame of reference rotating with the horizon
angular velocity 
$\Omega_{_H}$ 
coincides with the FFO frame on
the horizon. Therefore, the boundary conditions in this
horizon frame, which below is marked by a tilde,
\footnote{A tilde denotes the electromagnetic field components in a stationary
frame of reference that passes to the horizon reference frame,
i.e., to the frame rotating with the FFO angular velocity on the
horizon, 
$\Omega_{_H}={a\over r_{_H}r_g}$.}
are the same as (5):
$$
{\tilde F}_{\theta 0}\to 0\, ;\quad {\tilde F}_{\varphi 0}\to 0\, 
;\quad
{\tilde F}^{r\theta}\to 0\, ;\quad {\tilde F}^{r\varphi}\to 0\, .
\eqno(10)$$

\section{MAXWELL EQUATIONS}

Suppose that the current has a density with a deltashaped
distribution function in the meridional plane:
$$
j^i(r,\theta )=J^\varphi \delta^i_\varphi 
\left[ \delta (r-r_{_0})
\delta (\theta -\theta_{_0})\right] /\sqrt{-g} .
\eqno(11)$$
Since the other currents are assumed to be negligible
and since the frame is stationary, the toroidal electromagnetic-
field components are zero. Therefore, the axisymmetric
Maxwell equations for the covariant field
components outside the horizon are
$$
\left\{
\begin{array}{lll}
e^{\alpha\beta\varphi}\partial_\beta F_{\alpha\varphi}=0\, ;\\
e^{\alpha\beta\varphi}\partial_\beta F_{\alpha 0}=0\, .
\end{array}
\right.
\eqno(12)$$
We thus see that the covariant tensor components of the
electromagnetic field can be represented as
$$
F_{\alpha\varphi}=\partial_\alpha A_{\varphi}\, ;\qquad
F_{\alpha 0}=\partial_\alpha A_{_0}\, .
\eqno(13)$$
Here, 
$A_{i}$ ---
are the covariant components of the 4-vector
electromagnetic-field components. Let us now write the
axial Maxwell equations for the contravariant field
components without electric charges:
$$
\left\{
\begin{array}{rcl}
{1\over \sqrt{-g}}\partial_\alpha (\sqrt{-g}F^{\alpha 0})=0\, ,\\
{1\over\sqrt{-g}}\partial_\alpha (\sqrt{-g}F^{\alpha\varphi}) =
4\pi j^\varphi\, .
\end{array}\right.
\eqno(14)$$
The right-hand part of the second equation in (14) is
given by expression (11).

\section{DETERMINING THE MAGNETIC-FIELD
COMPONENTS}

Let us determine the magnetic field in the Schwarzschild metric 
($a=0$
in [4]). Because of axial symmetry,
the vector potential of the magnetic field has the toroidal
component alone. According to (13), we have for its
covariant part
$$
F^{\gamma\varphi}=
g^{\gamma\beta}g^{\varphi\varphi}\partial_\beta A_\varphi .
\eqno(15)$$
Denoting the tensor components in Euclidean space by
the subscript 
${_0}$, 
we then have, according to (6),
$$
F^{\gamma\varphi}= g_{_0}^{\gamma\beta}
(\delta^\gamma_\beta - {r_{_g}\over r}\delta^r_\beta \delta^\gamma_r )
 g_{_0}^{\varphi\varphi}\partial_\beta A_\varphi =
F_{_0}^{\beta\varphi}(\delta^\gamma_\beta -
{r_{_g}\over r}\delta^r_\beta \delta^\gamma_r ).
\eqno(16)$$
We see from (15) and (16) that if the function Aj is
smooth and has no singularities, then the boundary
conditions (5) are satisfied with the required asymptotics
(2). The second equation in (14) can then be
rewritten as
$$
{1\over\sqrt{-g}}\partial_\alpha (\sqrt{-g}F_{_0}^{\alpha\varphi})=
4\pi j^\varphi_{tot} .
\eqno(17)$$
where we designated
$j^\varphi_{tot}=j^\varphi +{r_{_g}\over 4\pi r^2}\partial_r
(rF_{_0}^{r\varphi})$.
Next, let us introduce the physical vector components
by the definition (see [3])
$$
\hat B^\alpha =B^\beta\sqrt{|g_{\alpha\beta}|}
\eqno(18)$$
In these components, Eq. (17) has the form of the Poisson
equation in Euclidean space for the vector potential
$\hat A^\gamma$
with source 
$\hat j^\gamma_{tot}$.
The solution of this equation is
known to be (see [3])
$$
\hat A^\alpha {_{(r,\theta )}}=\delta^\alpha_\varphi\hat e^\varphi
\displaystyle{\int\limits_{}^{}}
{{\hat {j'}}^\gamma_{tot} \hat e_\gamma\over |\vec r-\vec{r'}|}\, 
r'^2\, dr'\mathop{|}\limits_{r_g}^\infty \,\sin\theta'\, 
d\theta' \mathop{|}\limits_0^\pi\,
d\varphi' \mathop{|}\limits_{-\pi}^{\pi} .
\eqno(19)$$
Here, 
$\hat e^\gamma$ ---
is a unit vector
\footnote{${\hat {j'}}^\gamma_{tot} \hat e_\gamma
={j'}^\varphi_{tot} r'\sin\theta'\cos\varphi'$.}
in the direction of angle $\varphi$,
$|\vec r-\vec{r'}|^2=r^2+r'^2-2rr'(\cos\theta\cos\theta'+
\sin\theta\sin\theta'\cos\varphi' ),$
$\theta$ ---
is the inclination of vector $\bf r$ to the $z$ axis, 
$\theta'$ is the
inclination of vector $\bf r'$ to the $z$ axis, and 
$\varphi'$ is the angle
between the projections of vectors $\bf r$ and $\bf r'$ onto the
plane perpendicular to the $z$ axis. The integration is performed
in Euclidean space outside the sphere of radius $r_g$.
Solution (19) is obtained by iterations; we assume that
$r_{_g}=0$
in the initial iteration.\\
Let us determine the magnetic vector
$$
H_\lambda =-{\sqrt{-g}\over 2}e_{\lambda
\beta\gamma}F^{\beta\gamma} ,
\eqno(20a)$$
or
$$
F^{\alpha\beta}=-{1\over\sqrt{-g}}H_\lambda e^{\lambda\alpha\beta}.
\eqno(20b)$$
According to (18) and (20), the components 
$A_\varphi$, $H_r$ 
, and
$H_\theta$
are given by
$$
A_\varphi = -r\sin\theta\hat A^\varphi ,\quad
H_r ={\partial_\theta (\sin\theta\hat A^\varphi )\over r\sin\theta} ,\quad
H_\theta /r=-{1-{r_{g}\over r}\over r}\partial_r (r\hat A^\varphi ).
\eqno(21)$$
For the initial iteration, according to (21), the following
expressions can be derived for the magnetic-
field components from (19) and because of the presence
of delta functions in (11):
$$
\left\{
\begin{array}{lll}
H_r^{^0} {_{(r,\theta )}} =\left[{J^\varphi\over r_{_0}}\right]
{\displaystyle \int\limits_{-\pi}^{\pi}} \left\{ {\cos\theta\cos^2\varphi' 
(1+x^2-x\sin\theta\cos\varphi')\over x\sin\theta 
(1+x^2-2x\sin\theta\cos\varphi')^{3/2}}\right\}\, 
d\varphi' \\
H_\theta^{^0} {_{(r,\theta )}}/r =\left[{J^\varphi\over r_{_0}}\right]
{\displaystyle \int\limits_{-\pi}^{\pi}} \left\{ {x\sin\theta\cos^2\varphi' -
\cos\varphi'\over x
(1+x^2-2x\sin\theta\cos\varphi' )^{3/2}}\right\} \, d\varphi' .
\end{array}
\right.
\eqno(22)$$
Here, we designated 
$x=r/r_{_0}$.
In particular, we obtain
$$
\lim_{\theta\to 0}\, (H^{^0}_\theta /r) =0 ,\quad 
\lim_{x\to\infty}\, (H^{^0}_\theta /r) =
{\pi J^\varphi\sin\theta\over r_{_0}x^3}\, ,\quad 
\lim_{\theta\to 0}\, H^{^0}_r =
{2\pi J^\varphi\cos\theta\over r_{_0}(1+x^2)^{3/2}}.
\eqno(23)$$

\section{DETERMINING THE ELECTRIC POTENTIAL}

To calculate the electric potential 
$A_{_0}$, 
we use the formula
$$
F^{\alpha 0}=\{ F_{\beta 0}g^{^{00}}+F_{\beta\varphi}g^{^{0\varphi}}\}
g^{^{\alpha\beta}}
\eqno(24)$$
However, the components 
$F^{\alpha 0}$
must be expressed in
terms of
$\tilde F_{\beta 0}$  
and 
$\tilde F^{\beta\varphi}$, 
because we know the boundary
conditions only for them in stationary frames of 
reference. For the poloidal electromagnetic-field components,
using formulas (6), (8), and (9), and the expression
$g^{^{\alpha i}}g_{_{\gamma i}}=
g^{^{\alpha\beta}}g_{_{\gamma\beta}}=
\delta^{^\alpha}_{_\gamma}$,
\footnote{
For $\alpha$ running the values of $r$ and $\theta$, 
we have $g_{0\alpha} = 0$ and $g^{0\alpha} = 0$.}
we obtain
$$
F^{\alpha 0}=L_f\left[ g^{^{00}}g^{^{\alpha\beta}}\tilde A_0,{_\beta} +
\tilde F^{\alpha\varphi}(\Omega_{_H}-\Omega )/(\kappa +\Omega^2)\right] ,
\eqno(25)$$
where 
$L_f = \left[ 1-(\Omega -\Omega_{_H})^2/(\kappa+\Omega^2)\right]^{-1}$
has the meaning
of FFO Lorentz factor relative to the stationary frame of
the horizon.
Hence, in the second approximation in 
 $a^*=a/M=L/M^2$, 
the dimensionless black-hole angular momentum
($0<a^*<1$), the first equation in (14) can be reduced to
$$
\Delta\tilde A_{_0}=-4\pi\left( f_{(H_{\alpha})}
+U_{(\tilde A_{_0})}\right) ,
\eqno(26)$$
Here,
 $\Delta$ ---
 is the Laplace operator in Euclidean space and
$H_\alpha$ ---
are the Schwarzschild magnetic-field components
derived in the preceding section:\\
$\quad U_{(\tilde A_{_0})}~=~-{r_{_g}\over 
4\pi r^3}\partial_r(r^2\partial_r\tilde A_{_0}),$\quad
$f_{(H_{\alpha})}= {a\over 4\pi r^2_{_g}}(1-{r_{_g}\over r})\times \\
\times\left( (1+{r_{_g}\over r}+{r^2_{_g}\over r^2})[2\cos\theta H_{r} 
-4\pi j^\varphi r^2\sin^2\theta ]-
\sin\theta (2+{r_{_g}\over r})H_{\theta}/r \right) .$\\
The expression in round brackets on the right-hand
side of Eq. (26) is an analog of the electric charge density
in the Poisson equation. However, because of the
second term, the 
$\tilde A_{_0}$
---dependent function
$U$, 
it can be
solved by the iteration method assuming that 
$U=0$
in the initial iteration. According to the boundary conditions
(10), solving Eq. (26) is equivalent to calculating the potential
$\tilde A_{_0}$,
 produced by the density of electric charge 
$\rho =(f+U)$
around a conductive sphere of radius 
$r_{_g}$
in Euclidean space. This external problem for the Poisson
equation can be solved by the image method [6].
The solution that satisfies the boundary conditions (10) is
$$
\tilde A_{_0}{_{(r,\theta )}}=\int\limits^{}_{}\, \rho_{(r',\theta' )}\left\{
{1\over |\vec r-\vec{r'}|}-
{r_{_g}/r'\over |\vec r-\vec{r'}(r^2_{_g}/{r'}^2)|}\right\}\, 
r'^2\, dr'\mathop{|}\limits_{r_g}^\infty \,\sin\theta'\, 
d\theta' \mathop{|}\limits_0^\pi\,
d\varphi' \mathop{|}\limits_{-\pi}^{\pi}  .
\eqno(27)$$
The integration is performed in Euclidean space outside
the sphere of radius
$r_{_g}$
[see (19)]. According to (7), we obtain the potential 
$A_{_0}$
from (27)
$$
A_{_0}=\tilde A_{_0}-\Omega_{_H}\tilde A_\varphi \, ;
\quad A_\varphi =\tilde A_\varphi\, .
\eqno(28)$$
We can factor 
$a J^\varphi /r_{_0}$
outside integral (27). Denoting the
magnetic-field strength at the center of the system in the
absence of a black hole by 
$H_{_0}=2\pi J^\varphi /r_{_0}$ 
 [see (23)], we
derive for
$\tilde A_{_0}$
 and
$A_{_0}$:
$$
\tilde A_{_0} =\tilde{Int}\left({r\over r_{_g}},\theta\right)\cdot 
a^*\cdot M\cdot H_{_0}/(2\pi )\, ;\,\, \makebox{ è }\,\,
A_{_0} =Int\left({r\over r_{_g}},\theta\right)\cdot 
a^*\cdot M\cdot H_{_0}/(2\pi ).
\eqno(29)$$
Here, 
$Int\left({r\over r_{_g}},\theta\right)$
and 
$\tilde{Int}\left({r\over r_{_g}},\theta\right)$
are dimensionless functions,
which can be numerically calculated using expressions
(27) and (28). The corresponding results are
shown in the figure. Some important characteristics of
the solution can also be established analytically.

\begin{figure}[t]
\centering
\epsfbox[15 280 520 700]{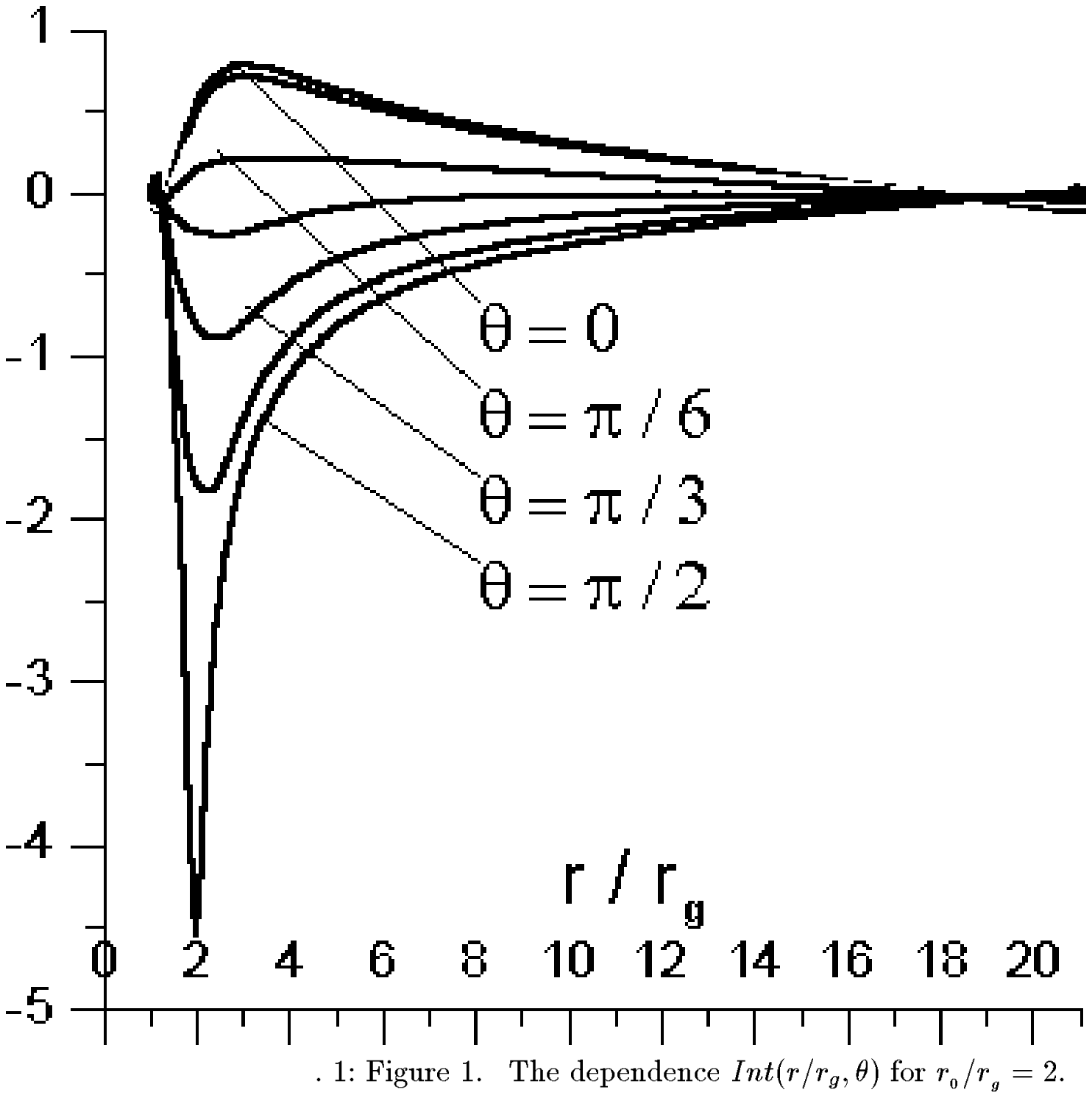}
\label{ris1}
\end{figure}

1). Since the system has a mirror symmetry relative
to the inversion in the equatorial plane,
\footnote{This can be seen from (26) and from the expression for 
$f$
($\sin\theta$ 
and 
$H_{\theta}$
are even, while 
$\cos\theta$
and 
$H_{r}$
are odd relative to the mirror
inversion).}
the electric-field direction on the axis depends on the coincidence
of the directions of black-hole angular momentum and
ring-current moment.

2). We see from (27) that 
 $\tilde A_{_0}\to 0$
 when 
$r\to r_{_H}$
(the expression in curly braces becomes zero). This
important result has a simple explanation:\\
Since the total charge under the horizon is zero, the
electric field must pierce the horizon in different directions.
Therefore, there must be an equipotential surface
that separates these directions, pierces the horizon, and
goes to infinity. Since the potential
$\tilde A_{_0}$
 changes neither
on this surface, nor on the horizon (in view of the
boundary conditions), the horizon potential is equal to
the potential at infinity (zero). This conclusion has an
important implication: the existence of a local extremum
for the potential
$\tilde A_{_0}$. 
Similar reasoning also applies to 
$A_{_0}$. 
It might seem that a nonzero electric
charge density must arise near the extremum. In reality,
however, this is not the case: the charge density is determined
by the contravariant electric-field components
($F^{\alpha 0}$ 
rather than 
$F_{\alpha 0}$),
while the derivation of (27) is
based on the first equation in (14), a zero charge density
in the entire space. We emphasize that there is no extremum
in the well-known solution of Wald [7] for a black
hole placed in a uniform magnetic field aligned along
the symmetry axis of the black hole. This is because the
field does not vanish at infinity in this solution, and
this nonphysical boundary condition wipes out the
extremum.

\section{THE KINETIC ENERGY OF A CHARGED
PARTICLE EMERGING ALONG THE AXIS}

The conserved mass–energy of a charged particle
with a zero angular momentum component along the
$z$ 
axis is given by [4, 5]
$$
E =m_{_0}\gamma_{_L}
\sqrt{{\rho^2{\scriptstyle\Delta}\over (r^2+a^2)^2-a^2{\scriptstyle\Delta}
\sin^2\theta}}-qA_{_0}=m_{_0}{\gamma_{_L}}^\infty ,
\eqno(30)$$
where 
$\gamma_{_L}$ --- 
 is the Lorentz factor of the particle, and 
$q$ --- 
is its charge.
Consider the acceleration of an emerging charged
particle along the  axis by an electric field. Since 
$qA_{_0}=-|q\tilde A_{_0}|$
in this case, the kinetic energy of this particle at
infinity with 
$\gamma_{_L}$, $g_{_{00}}$, and $A_{_0}$
specified at any point of the
$z$ 
axis is
$$
E_k  = E-m_{_0}=m_{_0}(\sqrt{g_{_{00}}}\gamma_{_L}-1)+|qA_{_0}| .
\eqno(31)$$
Since the electromagnetic energy of a charged particle
for quasars is much larger than the corresponding rest
energy, it is convenient to represent the result for 
$A_{_0}$
as the energy (in electronvolts) that an elementary charge
emerging along the axis acquires. For magnetic fields
$H_{_0}\approx 10^4$
and mass 
$M\approx 10^8\, M_{^{_\bigodot}}$
we obtain from (29)
and (31):\\
$E_k~\approx~a^*\cdot 10^{19} \cdot
Int\left({r\over r_{_g}}, 0\right)$ [eV].

\section{CONCLUSION}

We have calculated the electric field generated by a
rotating black hole that interacts with an external magnetic
field. As can be seen from the above discussion,
this field is quadrupole in nature, being actually an
analog of the pulsar electric field (unipolar inductor
model [8]).
The situation on the 
$z$
axis was not chosen by
chance. Because of strong magnetic fields, the Larmor
radii of a charged particle must be of the order of the
gravitational radius of the system, and the particle
acceleration mechanism will be effective only in directions
close to the 
$z$
axis, where the Lorentz force does
not act.
The reader may ask a legitimate question: Will a
strong electric field produce electron–positron
plasma near the horizon and will it destroy the force field 
$({\bf E\cdot H})~\ne~0$? 
At large gradients in 
$A_{_0}$, 
this can actually
happen, and the problem should then be solved in
the force-free approximation (see review papers [9–11]
on this subject). However, as numerical estimates show,
the electric-field strength in the model does not exceed
$10^7 \, V\, cm^{-1}$
while a strength of the order of 
 $\sim 10^9  \, V\, cm^{-1}$
is required for the particle production.
\footnote{The production of particles by inverse-Compton-type effects is
not considered, because the particle number density around the
black hole is assumed to be low.}
 In addition, in
the force-free approximation, particles cannot be accelerated
along the 
$z$ 
axis, because there is no electric field
in this direction. Besides, in any direction in a forcefree
field, the electric and magnetic fields equally act on
the particle. Therefore, even if its trajectory goes to
infinity, the particle loses most of its energy in the process.
Recall that jets are observed in quasars precisely
in the directions along and opposite to the axis. For this
reason, solving the problem under consideration in the
force-field approximation is of considerable importance.

$$ $$

ACKNOWLEDGMENTS

      I wish to thank N.S. Kardashev, V.S. Beskin,
V.N. Lukash, B.V. Komberg, M.I. Zel'nikov, Yu.Yu. Kovalev,
and all the remaining staff of the theoretical
departments at the Astrospace Center and the Lebedev
Institute of Physics, as well as participants of workshops
for taking an active part in preparing the article
and for important remarks. 

This study was supported by
the Russian Foundation for Basic Research (project
N 00-15-96698 and 01-02-17829).

$$ $$

REFERENCES

1. R. D. Blandford and R. L. Znajek, Mon. Not. R. Astron.
Soc. 179, 433 (1977).

2. J. M. Bardeen and J. A. Petterson, Astrophys. J. Lett.
195, L65 (1975).

3. L. D. Landau and E. M. Lifshitz, The Classical Theory
of Fields (Pergamon, Oxford, 1975; Nauka, Moscow,
1988).

4. Black Holes: the Membrane Paradigm, Ed. by
K. S. Thorne, R. H. Price, and D. A. Macdonald (Yale
Univ. Press, New Haven, 1986; Mir, Moscow, 1988).

5. I. D. Novikov and V. P. Frolov, Physics of Black Holes
(Nauka, Moscow, 1986).

6. L. D. Landau and E. M. Lifshitz, Course of Theoretical
Physics, Vol. 8: Electrodynamics of Continuous Media
(Nauka, Moscow, 1982; Pergamon, New York, 1984).

7. R. M. Wald, Phys. Rev. D 10, 1680 (1974).

8. P. Goldreich and W. Julian, Astrophys. J. 157, 869
(1969).

9. V. S. Beskin, Usp. Fiz. Nauk 167, 689 (1997) [Phys.
Usp. 40, 659 (1997)].

10. V. S. Beskin, Ya. N. Istomin, and V. I. Par'ev, Astron. Zh.
69, 1258 (1992) [Sov. Astron. 36, 642 (1992)].

11. J. S. Heyl, Phys. Rev. D 63, 064028 (2001); grqc/
0012007; A. Tomimatsu and H. Koyama, Phys. Rev. D
61, 124010 (2000); gr-qc/0002020.

Translated by V. Astakhov

\end{document}